\documentclass[preprint,showpacs,preprintnumbers,nofootinbib,amsmath,amssymb,aps,prc,longbibliography]{revtex4-1}

\usepackage{graphicx}
\usepackage{color}
\usepackage{longtable}
\usepackage{multirow}
\begin{document}


\title{Level scheme of $^{153}$Sm obtained from $^{152}$Sm($n_{th}$,$\gamma$) reaction using $\gamma-\gamma$ coincidence spectrometer}


\author{Nguyen Ngoc Anh$^1$}
\email[]{ngocanh8999@gmail.com}
\author{Nguyen Quang Hung$^2$}
\email[]{nguyenquanghung5@duytan.edu.vn}
\author{Nguyen Xuan Hai$^1$}
\email[]{xuannguyennri@gmail.com}
\author{Pham Dinh Khang$^3$}
\author{A. M. Sukhovoj$^4$}
\author{L. V. Mitsyna$^4$}
\author{Ho Huu Thang$^1$}
\author{Le Hong Khiem$^{5,6}$}

\affiliation{1) Dalat Nuclear Research Institute, Vietnam Atomic Energy Institute, 01 Nguyen Tu Luc, Dalat City, Vietnam \\ 
2) Institute of Fundamental and Applied Sciences, Duy Tan University, Ho Chi Minh City 700000, Vietnam \\
3) Hanoi University of Science and Technology, 01 Dai Co Viet, Hanoi City, Vietnam \\
4) Joint Institute for Nuclear Research, Dubna, Russia \\
5) Institute of Physics, Vietnam Academy of Science and Technology, Hanoi City, Vietnam \\
6) Graduate University of Science and Technology, Vietnam Academy of Science and Technology, Hanoi City, Vietnam}


\date{\today}

\begin{abstract}

The level scheme of the compound $^{153}$Sm nucleus formed via the $^{152}$Sm($n_{th}$,$\gamma$) reaction is studied by using the $\gamma-\gamma$ coincidence spectrometer at Dalat Nuclear Research Institute, Vietnam. All the gamma cascades, which correspond to the decays from the compound state to 12 final levels of 0 ($\frac{3}{2}^+$), 7.535 ($\frac{5}{2}^+$), 35.844 ($\frac{3}{2}^-$), 90.875 ($\frac{5}{2}^-$), 126.412 ($\frac{1}{2}^-$), 127.298 ($\frac{3}{2}^-$), 182.902 ($\frac{5}{2}^-$), 321.113 ($\frac{3}{2}^+$), 404.129 ($\frac{1}{2}^-$), 405.470 ($\frac{3}{2}^-$), 414.924 ($\frac{1}{2}^+$), and 481.088 ($\frac{3}{2}^+$) keV, have been measured. A total number of 386 cascades corresponding to 576 gamma transitions has been detected. Among these cascades, 103 primary gamma transitions together with their corresponding intermediate levels and 299 secondary transitions have been determined. In addition, 29 primary gamma transitions, 42 intermediate levels, and 8 secondary transitions have been found to be the same as those extracted from the ENSDF data. The remain 74 primary gamma transitions, 61 intermediate levels, and 291 secondary transitions are therefore considered as the new data. In particular, based on an assumption that most of the transitions are dipole, we have tentatively assigned the unique spin value of $\frac{3}{2}\hbar$ for 53 observed intermediate levels corresponding to the cascades from the compound state to the final ones of 7.535 ($\frac{5}{2}^+$), 90.875 ($\frac{5}{2}^-$), and 182.902 ($\frac{5}{2}^-$) keV, whereas the remain levels are assigned with the spin values in the range of $[\frac{1}{2},\frac{3}{2}]\hbar$. Moreover, the total and partial (for the spin range of $[\frac{1}{2},\frac{3}{2}]\hbar$) cumulative numbers of levels have been constructed by combining the ENSDF data with the new data obtained within the present experiment. Comparison between these new cumulative curves and those extracted from the nuclear level density (NLD) data, which are obtained by using the Oslo method, shows that the maximum excitation energy $E_{\rm max}$, defined as the energy threshold below which most of the excited levels are observed, is extended to about 1.2 and 1.8 MeV for the total and partial NLD data, respectively. These values of $E_{\rm max}$ are higher than those obtained by using the present ENSDF data, which are around 1 MeV. The new cumulative curves have also been compared with different phenomenological and microscopic NLD models and the recent exact pairing plus independent-particle model at finite temperature (EP+IPM), in which no fitting parameter has been employed, is found to be the best fitted one. The present findings are very important as they will provide the updated information on the nuclear level structure and make a step forward to the completed level schemes of excited compound nuclei. 
\end{abstract}


\maketitle


\section{Introduction} \label{sec: intro}
The energy-level properties of excited nuclei (called the nuclear level scheme), which include the level energies, spins, parities, and gamma-rays associated with the excited levels, are important for the study of nuclear structure physics, nuclear reactions, and nuclear astrophysics. The level schemes of nuclei in the mass region $A\sim 150 \div 154$ are of particular interest because the nuclear deformation in this region was predicted to change drastically with only slight variations of $A$ \cite{1969Sm04,1979Re04}. Nuclei in this mass region are also called transitional nuclei. For example, $^{150}$Sm and $^{152}$Sm have very different level schemes as the former has the vibrational/quasi-vibrational characteristics, whereas the latter follows the rotational ones \cite{1968Lure}. Similarly, the level spectrum of $^{152}$Sm shows both rotational and vibrational behaviors, whereas that of $^{154}$Sm exhibits the strong rotational properties, indicating that this nucleus is strongly deformed \cite{1964Robert}. Moreover, two odd nuclei, $^{151}$Sm and $^{153}$Sm, which fall, respectively, between the two sets ($^{150}$Sm, $^{152}$Sm) and ($^{152}$Sm, $^{154}$Sm) are expected to be affected by the interplay between the rotational and vibrational bands \cite{1971Be41}. Therefore, the level schemes of $^{151,153}$Sm odd nuclei have been an interesting subject of many experimental and theoretical studies. The present paper focuses on the experimental study of the level scheme of $^{153}$Sm by using thermal neutron-capture reaction.

The level scheme of $^{153}$Sm has been studied by using different nuclear reactions and techniques \cite{Helmer2006} and all the experimental data have been compiled in the ENSDF library \cite{ENSDF}. For instance, the level scheme of $^{153}$Sm at the low-energy(spin) region (below 1.53 MeV) was studied by using the $\beta^-$ decay of $^{153}$Pm as well as the decay from the isomeric state of $^{153}$Sm to its ground state \cite{1971KiZC,1983MaYP,1995Gr19,1997Gr09}. These experiments detected in total 25 excited levels, 17 of which have the unique spin values within the interval of $[\frac{1}{2},\frac{9}{2}]\hbar$. The high-spin part in the level scheme of $^{153}$Sm was measured by using the heavy-ion capture reactions, in which a total number of 28 excited levels, 25 of which have the unique spin values falling into the range of $[\frac{11}{2},\frac{41}{2}]\hbar$, was reported \cite{1979Re04,1999As05,2000Ha59}. However, the above experiments have not covered the excited levels, whose energy and spin are in the regions of $[1.5,4.0]$ MeV and $[\frac{1}{2},\frac{3}{2}]\hbar$, respectively. In these regions, several transfer reactions such as $^{151}(t, p)$ \cite{2005Bu21}, $^{152}$Sm$(d, p)$ \cite{1965Ke09,1972Ka07,1997GoZn},$^{154}$Sm$(d, t)$ \cite{1971Be41,1972Ka07,1997GoZn}, $^{154}$Sm$(p, d)$ \cite{1997GoZn, 1997Bl11}, $^{152}$Sm($\alpha,^{3}$He) \cite{1984Li02}, $^{154}$Sm($^3$He, $\alpha$) \cite{1997GoZn}, and $^{154}$Eu($t, \alpha$) \cite{1985Ma26} have been employed and a considerable number of excited levels of $^{153}$Sm within the spin range of $[\frac{1}{2},\frac{11}{2}]\hbar$ has been explored. Most importantly, by using the $^{152}$Sm$(d, p)$ reaction, 132 excited levels below 3.929 MeV and 56 excited levels below 1.991 MeV in the level scheme of $^{153}$Sm have been deduced in Refs. \cite{1965Ke09} and \cite{1972Ka07}, respectively. Although, the data reported in Refs. \cite{1965Ke09} and \cite{1972Ka07} agree with each other, their uncertainties are quite high (about 10 keV or higher). The reason is that within the framework of the transfer reactions, the excited levels are indirectly deduced from the energy and momentum distributions of the reaction products (charged particles), instead of the direct way, that is, from the gamma transitions of the excited levels. The latter were also not reported in Refs. \cite{1965Ke09} and \cite{1972Ka07}.

Apart from the above ion-induced experiments, the neutron-captured reactions also play an important role in the construction of the $^{153}$Sm level scheme. In fact, by using the ($n_{th},\gamma$) and ($n$ = 2 keV, $\gamma)$ reactions ($n_{th}$ means the thermal neutron with energy of 0.025 eV), Refs. \cite{1971Be41,1969Sm04,1969Re04,1997GoZn} have thoughtfully investigated the level scheme of $^{153}$Sm by means of the bent-crystal, conversion-electron, and Ge detector spectrometers. For the latter, the first two spectrometers, which were used to measure the low energy gamma-rays, focused on the low-energy part (below 0.4 MeV) of the $^{153}$Sm level scheme, whereas the last one was used to detect the high-energy gamma rays and to consequently deduce the feeding levels corresponding to the observed gamma rays. Moreover, through the gamma spectrum measured by the Ge detectors, 35 gamma rays emitted from the compound state of $^{153}$Sm via ($n_{th},\gamma$) reaction were reported in Refs. \cite{1971Be41,1969Sm04,1969Re04}. Similarly, Ref. \cite{1997GoZn} has detected 31 gamma rays via ($n$ = 2 keV, $\gamma)$ reaction. Many excited levels, whose energies range from 0 to approximately 2.7 MeV, were also deduced from the gamma rays detected in Ref. \cite{1997GoZn}. In general, the number of gamma rays that can be detected by the conventional Ge detector spectrometer is restricted by the high Compton background of the gamma spectrum as well as the energy resolution of the Ge detector. Besides, the gamma spectrum of $^{153}$Sm obtained from the ($n,\gamma$) reaction is always influenced by $^{150}$Sm because the thermal neutron-capture cross section of $^{149}$Sm is extremely higher than that of $^{152}$Sm (See e.g., Table \ref{tab1}). 

Given the limitations of the works mentioned above, it is necessary to improve the level scheme of $^{153}$Sm, especially in the energy region from 0.5 MeV to about 5.0 MeV.  One of the possibilities is to perform the $^{152}$Sm($n_{th},\gamma$) reaction using an advance $\gamma-\gamma$ coincidence technique together with the Ge(Li) detectors (also called the (n, 2$\gamma$) technique or the method of digital summation amplitudes of coincident pulses) \cite{boneva1991}. This technique, which has advantages in identifying the correlated gamma transitions and in subtracting most of the Compton background, allows us to detect the two-step gamma cascades (TSC) decayed from the compound state to the low-energy final levels and can therefore be used to deduce many new excited levels in $^{153}$Sm within the energy region from 0.5 MeV to approximately 5.0 MeV and the spin range of $[\frac{1}{2}, \frac{3}{2}]\hbar$. Indeed, by using the above technique, we have successfully studied the updated level scheme of $^{172}$Yb via $^{171}$Yb($n_{th},\gamma$) reaction \cite{anh2017}. In particular, we have detected in the level scheme of $^{172}$Yb several new excited levels and the corresponding gamma transitions, whose data do not currently exist in the ENSDF library, especially in the intermediate energy region from 3 to 5 MeV.

The goal of the present paper is to update the level scheme of $^{153}$Sm via the ($n_{th},\gamma$) reaction by using the $\gamma-\gamma$ coincidence technique. The energy and spin regions to be covered by this experiment are $[0.52,5.3]$ MeV and $[\frac{1}{2}, \frac{3}{2}]\hbar$, respectively. In addition, by combining our newly updated levels with those presently existed in the ENSDF library, we are able to construct the new total and partial (within spin range of $[\frac{1}{2}, \frac{3}{2}]\hbar$) cumulative numbers of discrete levels, which are latter used to test the predictive power of various nuclear level density (NLD) models. At the same time, these new cumulative curves have also been compared with those extracted from the NLD data obtained by using the Oslo method \cite{OsloMethod}.

\section{Experimental Method}

The $^{152}$Sm($n_{th},\gamma$) reaction was carried out at Dalat Nuclear Research Institute (Vietnam) using the thermal neutron beam from the tangential channel of Dalat Nuclear Research Reactor. The thermal neutron beam, which was obtained by using the filtered technique, has the size and flux at the irradiated position to be equal to 2.5 cm and 1.7 $\times$ 10$^5$ n.cm$^{-2}$.s$^{-1}$, respectively. This beam configuration is sufficient for the present experiment as discussed e.g., in Ref. \cite{anh2017}. The experimental setup and measurement using the $\gamma-\gamma$ coincidence spectrometer with two HPGe detectors are the same as those presented in Ref. \cite{anh2017} (except the target nucleus), so we do not repeat them here.

The target nucleus $^{152}$Sm is in the form of a 583 mg Sm$_2$O$_3$ powder. This target, which was put in a plastic bag, was then measured at the center of the thermal neutron beam during approximately 661 hours. The isotopic content of the target, which is provided by the JSC Isotope Supplier with the quality certificate being given under the Contract No. 704/08625142/25/30-16, together with the thermal neutron-capture cross sections ($\sigma_{th}$) of all the isotopic components \cite{ncs} are given in Table \ref{tab1}. 
 
 \begin{table}[h!]
 \caption{Isotopic content of the target used in the present experiment.}
\begin{tabular}{c| c| c}
Isotope & Percentage (\%) & $\sigma_{th}$ (barn) \cite{ncs}\\ \hline
$^{152}$Sm & 98.7 & 206 $\pm$ 3 \\
$^{144}$Sm & 0.01 & 1.64 $\pm$ 0.10 \\
$^{147}$Sm & 0.06 & 57 $\pm$ 3 \\
$^{148}$Sm & 0.07 & 2.4 $\pm$ 0.6 \\
$^{149}$Sm & 0.13 & 40140 $\pm$ 600 \\
$^{150}$Sm & 0.20 & 100 $\pm$ 4 \\
$^{154}$Sm & 0.83 & 8.5 $\pm$ 0.5 \\
\end{tabular}
\label{tab1}
\end{table}

\begin{figure}[h]
 \centering
 \includegraphics[width = 12.9cm]{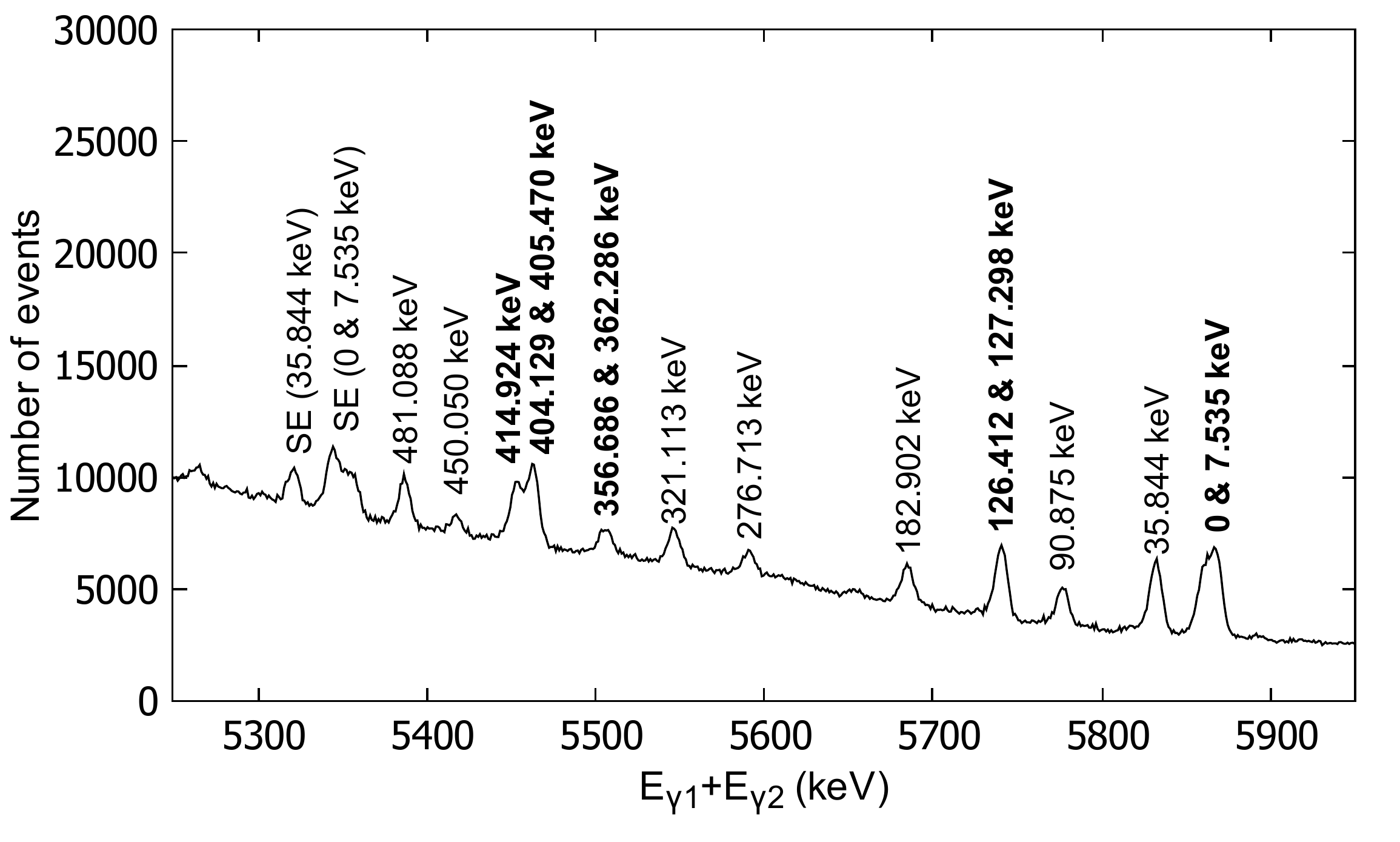}
 \caption{\label{sum153} Experimental summation spectrum of $^{153}$Sm. The final energies $E_f$ are marked on top of their corresponding peaks. The notation SE denotes the single-escape peaks.}
\end{figure}

Table \ref{tab1} shows that $^{144,148,154}$Sm isotopes have the values of both concentration and $\sigma_{th}$ being significantly smaller than those of $^{152}$Sm. Consequently, their influence on the spectroscopic data is negligible. For $^{147,150}$Sm isotopes, although their $\sigma_{th}$ values are comparable with that of $^{152}$Sm, their impact on the spectroscopic data is still small because of their tiny percentages. The only samarium isotope, which has a considerable influence on the spectroscopic data, is $^{149}$Sm because it has the noticeable $\sigma_{th}$ value, namely $\sigma_{th}$ of $^{149}$Sm is $\sim$ 198 times higher than that of $^{152}$Sm. Therefore, despite the percentage of $^{149}$Sm is $\sim$ 759 times less than that of $^{152}$Sm, its contribution to the coincidence events caused by the thermal neutron capture of $^{149}$Sm is only $\sim$ 3.8 times less than that of $^{152}$Sm, implying that approximately 20\% of all the detected coincidence events will be affected by the excited compound $^{150}$Sm nucleus. Fortunately, the two-step cascades caused by $^{150}$Sm can be distinguished from those of $^{153}$Sm by using the $\gamma-\gamma$ coincidence method because their summation energies (the total energy of two gamma rays) are different. For instance, the summation energies of the cascades of $^{150}$Sm detected within the present experiment range from $\sim$ 6.0 MeV to its neutron binding energy $B_n=$ 7.9867 MeV \cite{AME2016}, whereas those of $^{153}$Sm vary from $\sim$ 5.2 MeV to 5.87 MeV as clearly seen in Fig. \ref{sum153}.

For every detected coincident events, the energies absorbed by two HPGe detectors are recorded. The gamma cascades, which come from the decays of the compound state, go through different intermediate levels, and reach the ground state and some defined final levels, can be identified in the form of appropriate peaks appearing in the summation spectrum. The latter is obtained by counting the number of events per an interval of total energy absorbed by two HPGe detectors.

\begin{figure}[h]
 \centering
 \includegraphics[width=11cm]{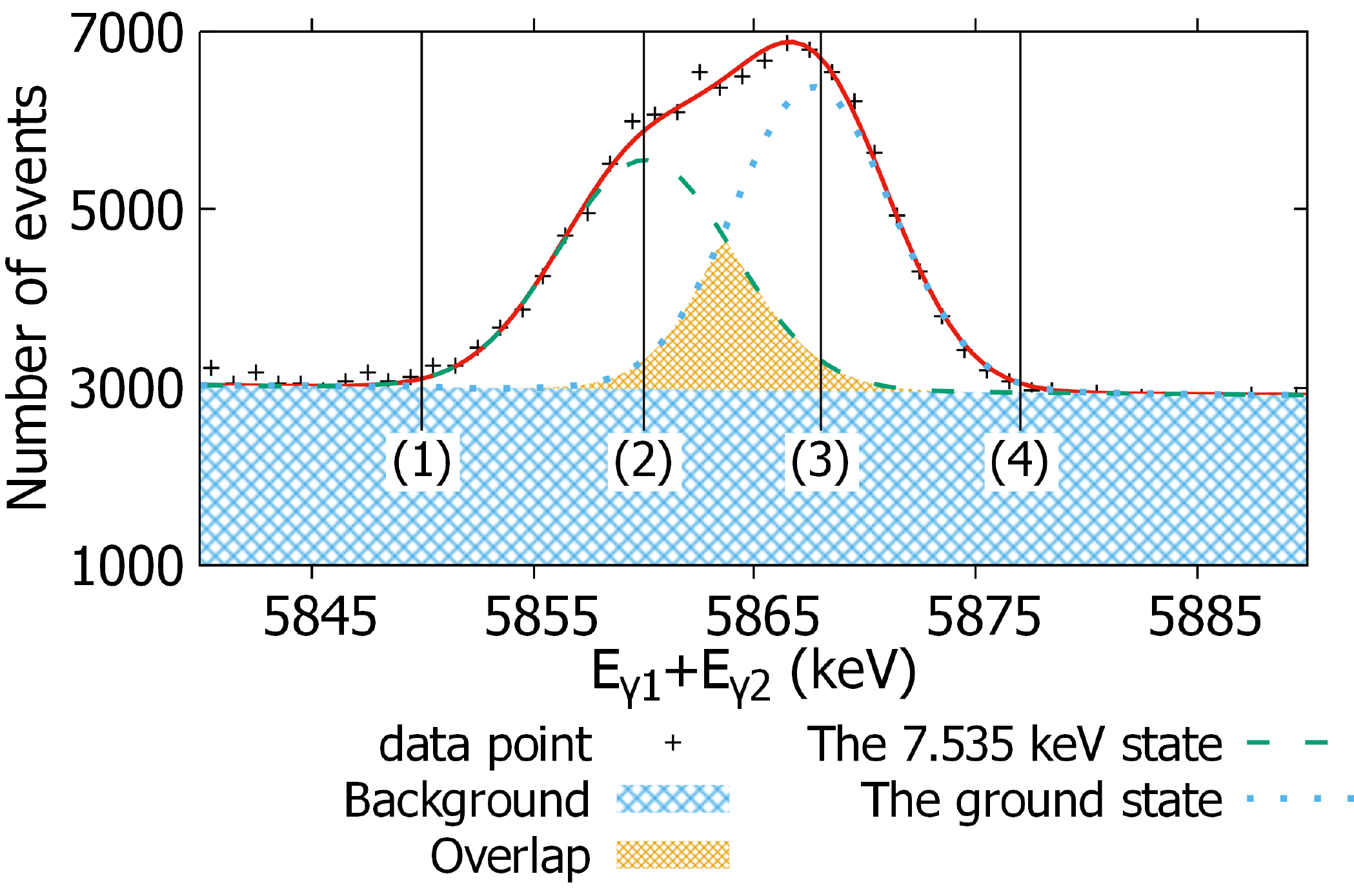}
 \caption{ \label{overlap} (Color online) Illustration of the gating windows used to reduce the contribution of the overlapped peaks. This figure shows the overlap of the summation peaks between the ground and 7.535 keV excited states.}
\end{figure}

The most instructive part of the summation spectrum of $^{153}$Sm is shown in Fig. \ref{sum153}. In this figure, all the gamma cascades decayed from the compound state to the ground state and 15 final states, whose energies are 7.535, 35.844, 90.875, 126.412, 127.298, 182.902, 276.713, 321.113, 356.686, 362.286, 404.129, 405.470, 414.924, 450.050 and 481.088 keV\footnote{It should be noted that the very precise energy values of the final levels given in the present paper are taken from Ref. \cite{Helmer2006}.}, can be identified based on their corresponding peaks. By gating on the appropriate peak, the TSC spectrum corresponding to the gamma cascades from the compound state to a given final level is obtained. Figure \ref{sum153} also shows some overlaps between different groups of states, whose energies are not much different, e.g. (0, 7.535 keV), (414.924, 404.129, and 405.470 keV), etc. The gamma cascades coming from these overlap peaks are indistinguishable because of the restricted energy resolution of the HPGe detectors used in the present experiment. However, these overlaps can be possibly reduced by a special selection of the gating window as illustrated in Fig. \ref{overlap}. It can be seen in this Fig. \ref{overlap} that an overlapped peak of two states can be fitted by two Gaussian functions, whose width and centroid position are different. Thus, the overlapped region can be easily identified if the gating window is divided into two regions. The first region is set between the lines (1) and (2) corresponding, respectively, to the head-tail and maximum positions of first Gaussian. The second region is chosen between the lines (3) and (4), which correspond to the maximum and end-tail positions of the second Gaussian, respectively. Once the overlapped region is identified (see the overlapped area in Fig. \ref{overlap}), its contribution can be easily reduced from the TSC spectrum. As a result, the contribution of the overlapped regions to the obtained TSC spectra is found to be less than 5\%. However, it should be noted that the above approach can not be applied if energies of the overlapped peaks are notably close to each others, namely the different between energies of two peaks is smaller than 0.8 FWHM (Full Width at Half Maximum), e.g. the following pairs of final levels (126.412, 127.298) keV and (404.129, 405.470) keV.

\begin{figure}[h]
 \centering
 \includegraphics[width=12.9cm]{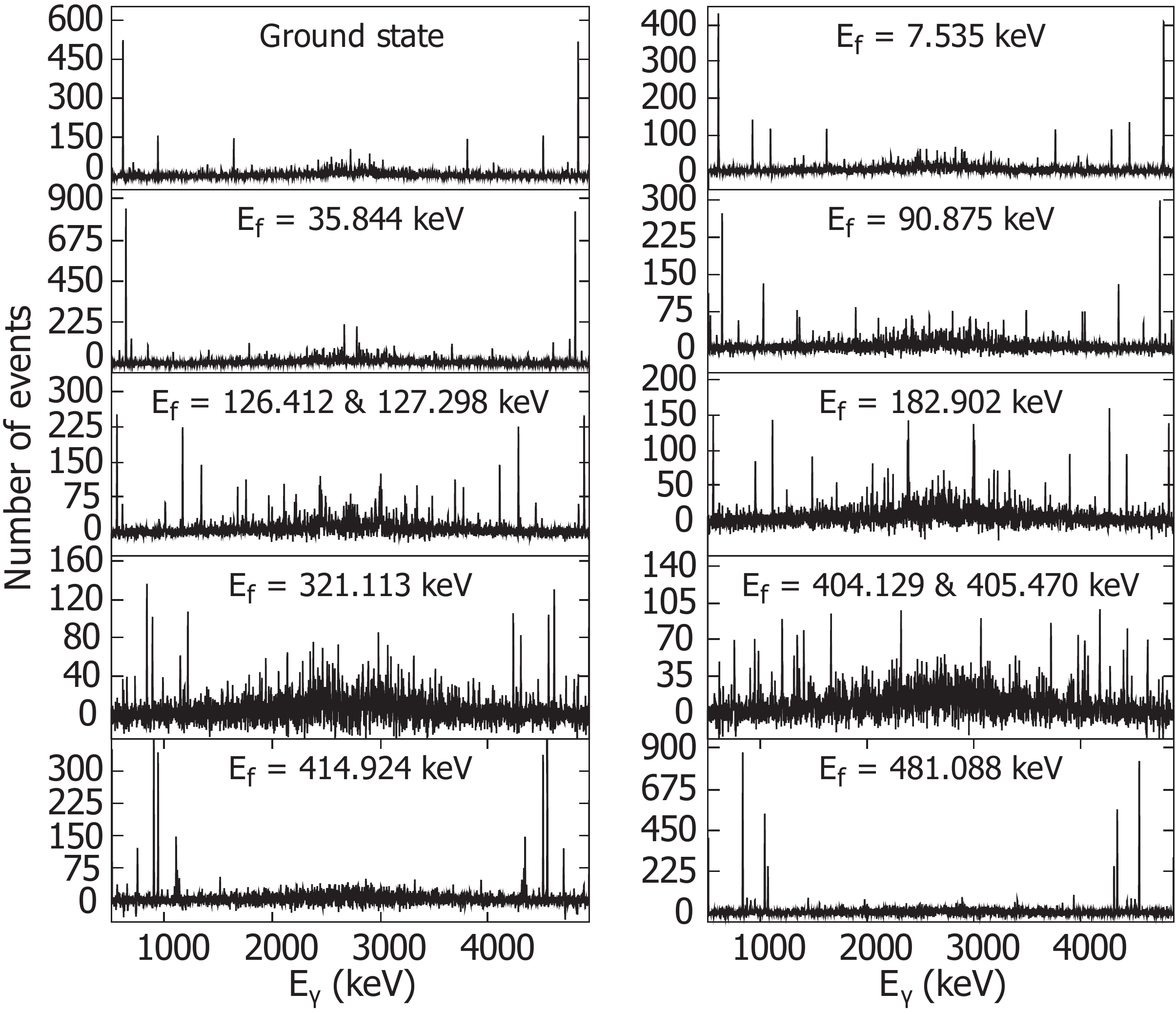} 
 \caption{Two-step cascade spectra of $^{153}$Sm obtained for different final states $E_f$.}
 \label{tsc}
\end{figure}

All the measured TSC spectra are shown in Fig. \ref{tsc}. Due to the low statistics, the TSC spectra corresponding to the following final levels 276.713, 356.686, 362.286, and 450.050 keV have not been analyzed yet. Despite the energy resolutions of the two HPGe detectors used in the present experiment are slightly different, the obtained TSC spectra are mirror symmetry because an algorithm for improving the digital resolution \cite{resolution} has been applied. The vicinity regions around each summation peak are gated to create a corresponding background spectrum. The latter is then subtracted from the spectrum obtained from the gating of the peak region, thus leading to some negative values in the TSC spectra in Fig. \ref{tsc}.

A pair of peaks, which are symmetric within a TSC spectrum, represents a gamma cascade. The peak positions and areas correspond to the transition energies and intensities, respectively. In order to construct the nuclear level scheme, we assume that the gamma transitions, which appear in more than one TSC spectrum, are considered to be the primary transitions. In addition, a transition is also considered as primary if it is currently determined as primary in the ENSDF library \cite{ENSDF}.

As for the spin of the levels, the possible spins of an observed intermediate level are often evaluated by using the following formula
\begin{equation}
\label{eq1}
\textrm{max}(J_i-L,J_f-L) \leq J \leq \textrm{min}(J_i+L,J_f+L),
\end{equation}
where $J_i, J$, and $J_f$ are spins of the initial, intermediate, and final levels, respectively, whereas $L$ is the multipolarity. Within the present work, we assume that all the observed transitions are dipole ($L=1$). This assumption is made because the probability of detecting the dipole transition is much higher than that of the quadrupole ($L=2$) \cite{blatt1991book}.
  
\section{Results and Discussion}
\subsection{Level scheme of $^{153}$Sm \label{nls}}
We have identified in total 576 gamma transitions corresponding to 386 gamma cascades, which are  associated with the decays from the compound state to the ground state and 11 final levels  (see Table \ref{tab2}). The latter are 7.535 ($\frac{5}{2}^+$), 35.844 ($\frac{3}{2}^-$), 90.875 ($\frac{5}{2}^-$), 126.412 ($\frac{1}{2}^-$), 127.298 ($\frac{3}{2}^-$), 182.902 ($\frac{5}{2}^-$), 321.113 ($\frac{3}{2}^+$), 404.129 ($\frac{1}{2}^-$), 405.470 ($\frac{3}{2}^-$), 414.924 ($\frac{1}{2}^+$), and 481.088 ($\frac{3}{2}^+$) keV. Based on these observed cascades, we have determined 103 primary gamma transitions corresponding to 103 intermediate levels and 299 secondary transitions emitted from these levels. Among the above primary transitions, 99 transitions have been deduced since they appear in more than one TSC spectrum. The remain 4 transitions, whose the energies are 4329.1, 4420.1, 4769.6, and 5133.2 keV, are also considered as the primary ones despite that they appear in only one TSC spectrum because these transitions are found to be the same as the primary transitions that currently exist in the ENSDF library \cite{Helmer2006}.

Since the compound state of $^{153}$Sm has the spin of $\frac{1}{2}\hbar$, by using Eq. (\ref{eq1}) together with an assumption that all the observed transitions are dipole, we are able to tentatively assign an unique spin value of $\frac{3}{2}\hbar$ for 53 intermediate levels, which correspond to the gamma transitions emitted from the compound state to 3 final levels with the spins of $\frac{5}{2}\hbar$, namely the 7.535 ($\frac{5}{2}^+$), 90.875 ($\frac{5}{2}^-$), and 182.902 ($\frac{5}{2}^-$) keV levels. For the remain 50 levels, which relate to the gamma transitions emitted from the compound state to the final levels with the spin of $\frac{1}{2}\hbar$ or $\frac{3}{2}\hbar$, their spin values can not be uniquely deduced. Consequently, a possible spin range from $\frac{1}{2}\hbar$ to $\frac{3}{2}\hbar$ has tentatively been assigned to these levels.

The assumption that all the observed transitions are dipole is made based on the following experimental evidences. First, among all the transitions coming from the compound state (see the ($n,\gamma$) datasets for thermal and 2-keV neutrons in Ref. \cite{Helmer2006}), we found only 2 transitions which are not dipole, namely the 5506.4 and 5861.4 keV transitions to the 362.286 ($\frac{5}{2}^+$) and 7.535 ($\frac{5}{2}^+$) keV levels, respectively. These transitions, however, have considerably low intensities compared to those obtained from other primary transitions. Moreover, the 5506.4 keV transition has solely found in Ref. \cite{1969Sm04}, whereas that of 5861.4 keV has only detected in the form of a doublet with the strong transition of 5868.4 keV in Refs. \cite{1969Sm04, 1969Re04, 1971Be41}, which has not been reproduced within the framework of ($n,\gamma$) experiment with 2-keV neutron \cite{1997GoZn}. Second, within the low-excitation energy of $^{153}$Sm level scheme, the quadrupole transitions have rarely been reported. In fact, there are only few quadrupole transitions, which currently existed in the ENSDF library, such as the 223.173 and 278.17 keV transitions coming from the 276.713 and 405.470 keV levels, respectively. They all together have lower energy than the energy threshold of the present work (520 keV for both transition and excitation energy). These  evidences apparently ensure the validity of the assumption above and consequently the reliability of the spin assignment within the present work, despite that the assumption is still restrictive and the spin assignment within the present work can not be determined as the  definite values.
 
By comparing the $^{153}$Sm level scheme obtained within the present work with that extracted from the ENSDF library \cite{Helmer2006}, we have realized that 29 primary gamma transitions and 42 intermediate levels are found to be the same within their uncertainties, whereas only 8 secondary transitions are the same with those existed in the ENSDF library. The remain 74 primary gamma transitions, 61 intermediate levels, and 291 secondary transitions are therefore considered as the new data obtained within the present experiment.

In particular, the $^{153}$Sm level scheme obtained within the present work agrees well with that obtained within the previous studies using the same $^{152}$Sm($n_{th},\gamma$) reaction \cite{1997GoZn,1971Be41,1969Sm04,1969Re04}. For the energy region below 5300 keV, which is the maximum gamma energy that can be detected within the present experiment (because the energy threshold of detectors were set to be around 520 keV), we have reproduced 19 over 24 primary transitions that were previously reported in Refs. \cite{1997GoZn,1971Be41,1969Sm04,1969Re04}. Among the 5 unreproduced transitions, 2 transitions, whose energies are 5220.4 and 5283.9 keV, were reported in Ref. \cite{1971Be41}, whereas 2 transitions with the energies of 4850 and 4864.0 keV were detected in Ref. \cite{1969Sm04}. These transitions were found very long time ago and have not been reproduced by other experiments. The remain 4505.6 keV transition was reported with a slightly different energy of 4506.6 $\pm$ 1.0 keV in Ref. \cite{1969Re04} or 4505.8 $\pm$ 0.4 keV in Ref. \cite{1971Be41}, or 4506.5 $\pm$ 0.6 keV in Ref. \cite{1969Sm04}. This 4505.6 keV transition might be therefore the same as the 4507.4 $\pm$ 0.4 keV transition observed within the present work as well as the 4507.41 keV transition obtained from the ($n,\gamma$) experiment with the 2-keV neutron source in Ref. \cite{1997GoZn}. In general, we have reproduced 22 over 26 levels that were reported by the previous ($n_{th},\gamma$) experiments within the excitation energy above 600 keV in Refs. \cite{1997GoZn,1971Be41,1969Sm04,1969Re04}.

The result of the $^{153}$Sm level scheme obtained within the present work also agrees well with the neutron capture experiment using 2-keV neutron source, namely 22 over 24 primary transitions within the gamma energy of 520 to 5300 keV and 23 over 29 levels within the excitation energy region of 600 to 2000 keV reported in Ref. \cite{1997GoZn} have been replicated within the present experiment. Among the remain unreproduced levels, 4 levels, whose energies are 1675.8, 1723.5, 1737.5, and 1751.4 keV, have been determined in Ref. \cite{1997GoZn} without any populating gamma transitions. In addition, all the levels reported in Ref. \cite{1997GoZn} with the assigned spins of $\frac{1}{2}\hbar$ or $\frac{3}{2}\hbar$ are fully in agreement with those deduced from the present study.

Furthermore, our data also go along with those obtained within the ion-induced experiments, in particular the  $^{152}$Sm($d,p$) \cite{1965Ke09,1972Ka07,1997GoZn}, $^{154}$($p,d$) \cite{1997GoZn,1997Bl11}, and $^{154}$($d,t$) \cite{1972Ka07,1997GoZn,1971Be41} reactions. Below 2000 keV, 24 excited levels found in the present work are supported by at least one of the experiments employing the ion-induced reactions. Similarly, 44 excited levels found within the present experiment agree with those extracted from the ion-induced reactions within their uncertainties (see the excited levels with the superscript denotation ''e'' in  Table \ref{tab2}). It should be noted here that the uncertainties of the data obtained within the ion-induced experiments are often in the range of 8 to 18 keV, which are much larger than those obtained within the present work. Therefore, we consider that two levels are the same only if their discrepancy is less than 1.5 keV, that is, if a level deduced from the present experiment agrees with that deduced from the ion-induced experiments but the discrepancy between the two levels is larger than 1.5 keV, it is considered as the new level.

\begin{longtable}{@{\extracolsep\fill}ccccc|p{2.5cm}ccc@{}}
\caption{Gamma-cascade transition energies and absolute intensities obtained from the $^{152}$Sm($n_{th},\gamma$) reaction. Primary transitions and intermediate levels corresponding to each gamma cascade are determined if possible. Comparisons between the data obtained within the present work with those extracted from the ENSDF library are made. Detailed explanation is given at the end of the table.}\\
\multicolumn{5}{c|}{Present Work}&\multicolumn{4}{c}{ENSDF}\\[0.05 cm] \hline
$E_1$ & $E_i$ & $J_i$ & $E_2$ & $I_{\gamma\gamma}$ & $E_f$ & $E_1$\textsuperscript{a} & $E_i$\textsuperscript{b} & $J_i$ \\
\hline
\endfirsthead
\caption[]{(continue)}\\
\multicolumn{5}{c|}{Present Work}&\multicolumn{4}{c}{ENSDF}\\[0.05 cm] \hline
$E_1$ & $E_i$ & $J_i$ & $E_2$ & $I_{\gamma\gamma}$ & $E_f$ & $E_1$\textsuperscript{a} & $E_i$\textsuperscript{b} & $J_i$ \\
\hline
\endhead
\textbf{5238.0 \textit{3}} & \textbf{630.4 \textit{7}} & $\mathbf{}\mathbf{\frac{3}{2}}$ & \textbf{630.4 \textit{3}} & {5039 \textit{327}} & 0 ($\frac{3}{2}^+$) & 5237.8 \textit{3} & 630.20 \textit{5}\textsuperscript{f} & $\frac{1}{2}^-,\frac{3}{2}^-$\\
 &  & $$ & \textbf{622.9 \textit{3}} & {3695 \textit{316}} & 7.535 ($\frac{5}{2}^+$) &  &  & \\
 &  & $$ & {594.6 \textit{4}} & {462 \textit{98}} & 35.844 ($\frac{3}{2}^-$) &  &  & \\
 &  & $$ & \textbf{539.5 \textit{3}} & {487 \textit{95}} & 90.875 ($\frac{5}{2}^-$) &  &  & \\\hline
\textsuperscript{x}{5231.2 \textit{4}} &  & $$ & \textsuperscript{x}{629.7 \textit{4}} & {485 \textit{114}} & 7.535 ($\frac{5}{2}^+$) &  &  & \\\hline
\textbf{5172.7 \textit{4}} & \textbf{695.7 \textit{8}} & $\mathbf{}{\frac{3}{2}}$ & {696.0 \textit{4}} & {312 \textit{77}} & 0 ($\frac{3}{2}^+$) & 5172.7 \textit{3} & 695.80 \textit{4}\textsuperscript{g} & $\frac{1}{2}^-$\\
 &  & $$ & \textbf{659.6 \textit{3}} & {4562 \textit{321}} & 35.844 ($\frac{3}{2}^-$) &  &  & \\
 &  & $$ & {604.8 \textit{5}} & {175 \textit{54}} & 90.875 ($\frac{5}{2}^-$) &  &  & \\
\end{longtable}

Table \ref{tab2} presents the absolute intensities normalized to 10$^6$ captures together with the statistical uncertainties of all 386 measured cascades. The normalization factor is determined based on the absolute intensities of 4697.2 and 5117.8 keV primary transitions (i.e., the 4697.4 and 5118.3 keV transitions within the present work) taken from the ENSDF data \cite{Helmer2006} together with their branching ratios. The latter are determined from the gating spectrum of the primary transitions mentioned above. Since the energy threshold of the present experiment is 520 keV, we are not able to identify the branches, whose energy of the secondary transition is less than 520 keV. Therefore, our cascade intensities may contain a certain systematic error.

In general, the present experiment reproduces most of the ENSDF data obtained from the neutron capture and ion-induced reactions. This consistency obviously proves the reliability of the data obtained within the present study.

Thanks to the coincidence technique, the influence of $^{150}$Sm on the spectroscopic information of $^{153}$Sm, which limits the number of data obtained from the neutron-capture experiment using the conventional HPGe detector \cite{1971Be41,1969Sm04, 1969Re04}, has been considerably reduced within the present experiment. This technique also reduces the peak overlaps, which are immensely common in analyzing the conventional prompt gamma spectra, especially for nuclei with the complicated level scheme such as in the case of $^{153}$Sm. The reason is that the coincidence technique is able to detect only the intermediate level in a narrow spin range from $J_i-1$ to $J_i+1$ ($J_i$ is the spin of the compound state) and the detected gamma transitions are distributed to the multiple TSC spectra. As a result, we are able to detect more important information on the level scheme of $^{153}$Sm, which have not currently existed in the ENSDF library. 

\subsection{Cumulative number of levels}
\subsubsection{Experimental cumulative number of levels} \label{cump1}
 \begin{figure}[h]
 \includegraphics[width=17.2cm]{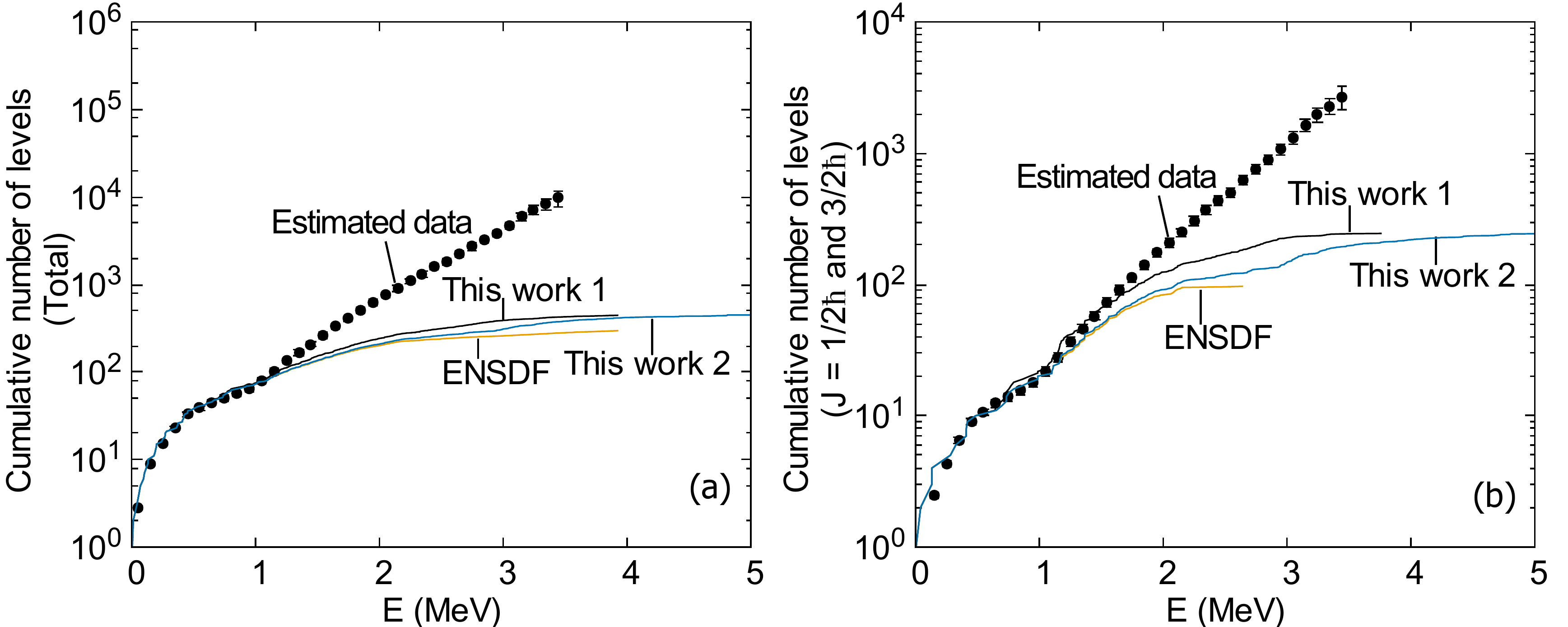}
 \caption{\label{cul} (Color online) Total (a) and partial (b) cumulative numbers of levels obtained by using the NLD data in Ref. \cite{Oslo} (estimated data) and ENSDF data in Ref. \cite{Helmer2006} in comparison with those obtained from ``This work 1'' and ``This work 2" (see the explanation in the text).}
 \end{figure}
 \begin{figure}[h]
 \includegraphics[width=12cm]{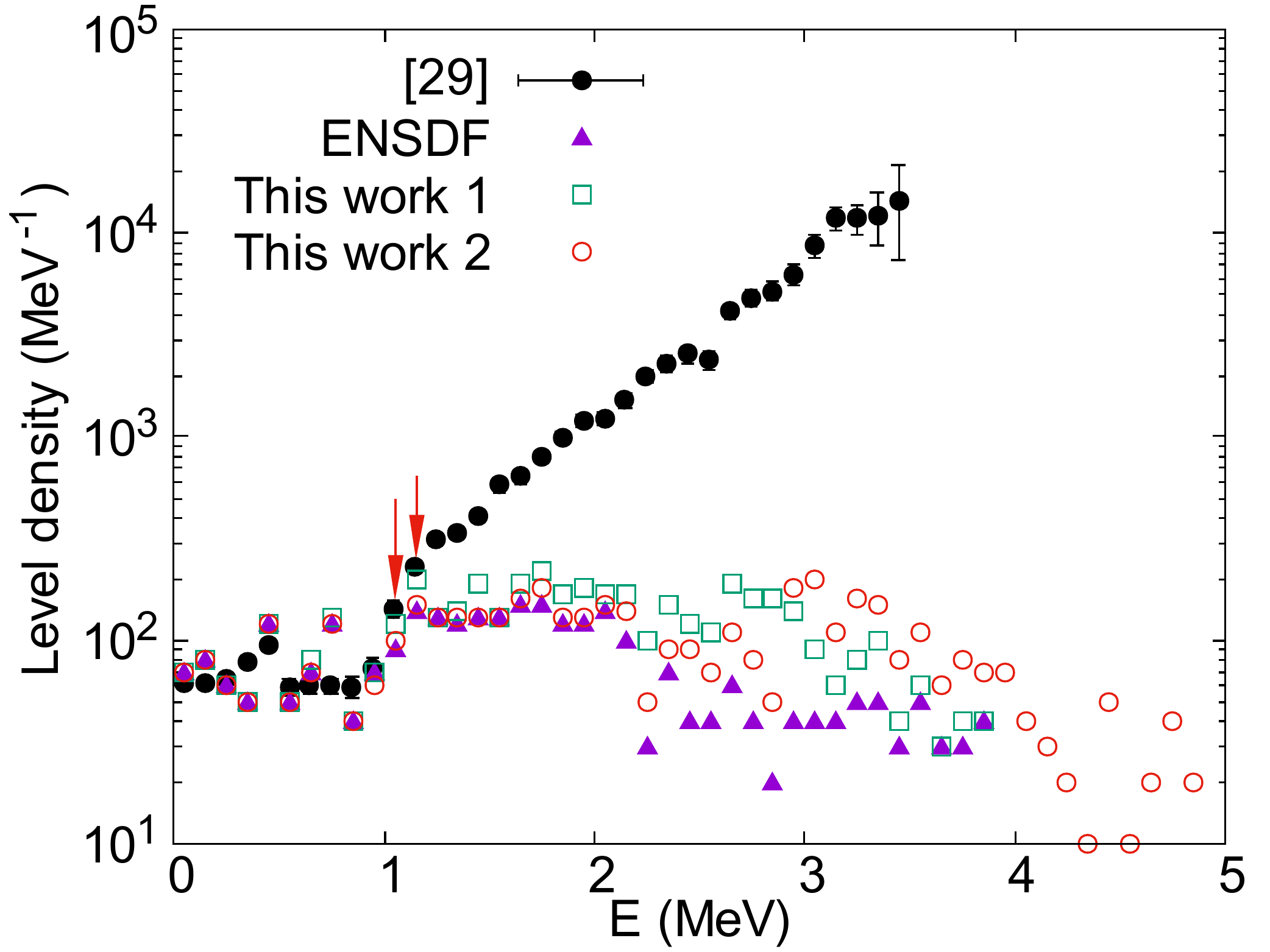}
 \caption{\label{nld} (Color online) Total level density obtained by counting the numbers of discrete levels in the ENSDF, ``This work 1'' and ``This work 2" versus the NLD data taken from Ref. \cite{Oslo}.}
 \end{figure}

Since several new energy levels have been detected within the present experiment, we are able to construct the total and partial cumulative numbers of levels, which are, by definition, the numbers of excited levels fall within the specific energy and spin ranges. These cumulative numbers are constructed by combining the adopted levels taken from the ENSDF \cite{Helmer2006} with those obtained within the present work (Table \ref{tab2}). For the latter, however, there are unassigned intermediate levels corresponding to 87 gamma cascades as shown in Table \ref{tab2} with the superscript denotation ''x''. Therefore, we have constructed two cumulative curves denoted by ``This work 1" and ``This work 2" (see Fig. \ref{cul}). ``This work 1" is created by assuming that the gamma transitions in each of 87 cascades with the higher energies are considered as the primary transitions, whereas those with lower energies correspond to the secondary ones. ``This work 2" is generated by using the opposite assumption, namely the gamma transitions with lower (higher) energies are considered as the primary (secondary) ones. It is obvious that ``This work 1" is always higher than ``This work 2", regardless of their total or partial cumulative curves because ``This work 1" contains the primary gamma transitions, whose energies are higher than those in ``This work 2" (Fig. \ref{cul}). Here, it should be noted that the assumption for ``This work 1" should be much more reliable than that for ``This work 2'' because within the two-step cascades, one often observes the primary transition, whose energy is higher than that of the secondary one (see e.g. the data reported in the ENSDF library \cite{ENSDF}). Consequently, the real cumulative curve should probably be very close to ``This work 1''. 

The total and partial cumulative numbers of levels within the present work are also compared with those obtained by using the NLD data in Ref. \cite{Oslo}. The total cumulative curve in this case is calculated by using the conventional formula \cite{gcmodel}
\begin{equation}
N(E_x) = \int_{0}^{E_x} \rho(E)dE ~, \label{ne_total}
\end{equation}
where $\rho(E)$ is the experimental NLD taken from Ref. \cite{Oslo}. As for the partial cumulative curve for the spin range $J= [\frac{1}{2},\frac{3}{2}]\hbar$, it should be calculated using the same Eq. (\ref{ne_total}) but the $J$-dependent NLD $\rho(E,J)$ must be used instead of the total NLD $\rho(E)$. However, there exists in literature only the total NLD extracted by using the Oslo method $\rho(E)$ in Ref. \cite{Oslo}. The latter was extracted from the gamma spectra of the $^{154}$Sm$(p, d\gamma)^{153}$Sm reaction, which were later normalized using the discrete levels taken from the ENSDF library \cite{ENSDF} as well as the NLD data at the neutron binding energy (see e.g., Fig. 3 of Ref. \cite{Oslo}). Therefore, in order to estimate the $\rho(E,J)$ values, we have manually multiplied $\rho(E)$ with a factor, which is determined as the ratio between the number of levels with spins $J =\frac{1}{2}$ and $\frac{3}{2}\hbar$ and the total number of levels existed in the ENSDF library \cite{Helmer2006}. This factor is found to be about 0.27 for $^{153}$Sm. The obtained $\rho(E,J)$ is then used to calculate the partial cumulative curve $N(E_x,J)$ for $J= [\frac{1}{2}, \frac{3}{2}] \hbar$. For the sake of simplicity, the corresponding results, namely the total and partial cumulative curves estimated using the NLD data in Ref. \cite{Oslo}, are called the estimated data/curves hereafter. It is seen in Figs. \ref{cul}(a) and (b) that such an estimation seems to be valid for the low-energy region (below 1 MeV) as both estimated curves for the total and partial cumulative numbers of levels are in excellent agreement with the ENSDF data. It is obvious that the spin distribution is not constant over the excitation energy. Thus, the estimated data presented in Fig. \ref{cul}(b) may not be corrected in the high-energy region above 1 MeV. Since the spin distribution changes very slightly when the excitation energy is low, we believe that our deduction is acceptable with a negligible error for the energy region from 1 MeV to 2 MeV. It is interesting to see in Fig. \ref{cul}(b) that ``This work 1'' almost coincides with the estimated data in the energy region from 0 to about 1.8 MeV, above which the data obtained from our estimation might be no longer valid. ``This work 2'' and ENSDF curves agree with the estimated data up to about 1 MeV only. This result supports strongly the validity of the assumption for ``This work 1'', which is the most common assumption used in the two-step cascade experiments as explained above. This assumption can also be confirmed by comparing the total NLD in Ref. \cite{Oslo} with those obtained from the ENSDF, ``This work 1'', and ``This work 2'' (Fig. \ref{nld}). It is clearly to see in Fig. \ref{nld} that the total NLDs taken from the ENSDF and ``This work 2'' only agree with the data of Ref. \cite{Oslo} below 1 MeV, whereas the agreement between ``This work 1'' and Ref. \cite{Oslo}'s data is extended up to about 1.2 MeV, indicating by two arrows in Fig. \ref{nld}. 

The results obtained from "This work 1" as shown in Figs. 4 and 5 indicate two significant contributions of the new levels found within the present work. The first contribution is that for the total NLD, the maximum excitation energy $E_{\rm max}$, defined as the energy threshold below which most of the excited levels have been observed, is now extended to about 1.2 MeV, instead of 1.0 MeV as that obtained from the ENSDF data \cite{Helmer2006} (Figs. \ref{cul}(a) and \ref{nld}). The second contribution is associated with the value of $E_{\rm max}$ for the spin range of [$\frac{1}{2},\frac{3}{2}$]$\hbar$, which has been increased up to about 1.8 MeV (Fig. \ref{cul}(b)). It is evident that the NLD calculated by counting the numbers of discrete levels has been widely considered as the most reliable data, which are often used for the normalization of the experimentally extracted data \cite{OsloMethod} as well as different NLD model calculations \cite{hfbcs,hfb}. However, the present ENSDF library provides the reliable NLD up to about 1 MeV only. By including our new data, we are able to obtain, for the first time, the reliable NLD data up to about 1.2 MeV and 1.8 MeV for the total and partial (within the spin range of [$\frac{1}{2},\frac{3}{2}$]$\hbar$) NLDs, respectively. This second contribution is therefore the most important contribution of the present work. 
 
\subsubsection{Comparison with theoretical models}

\begin{figure}[h]
\includegraphics[width=16cm]{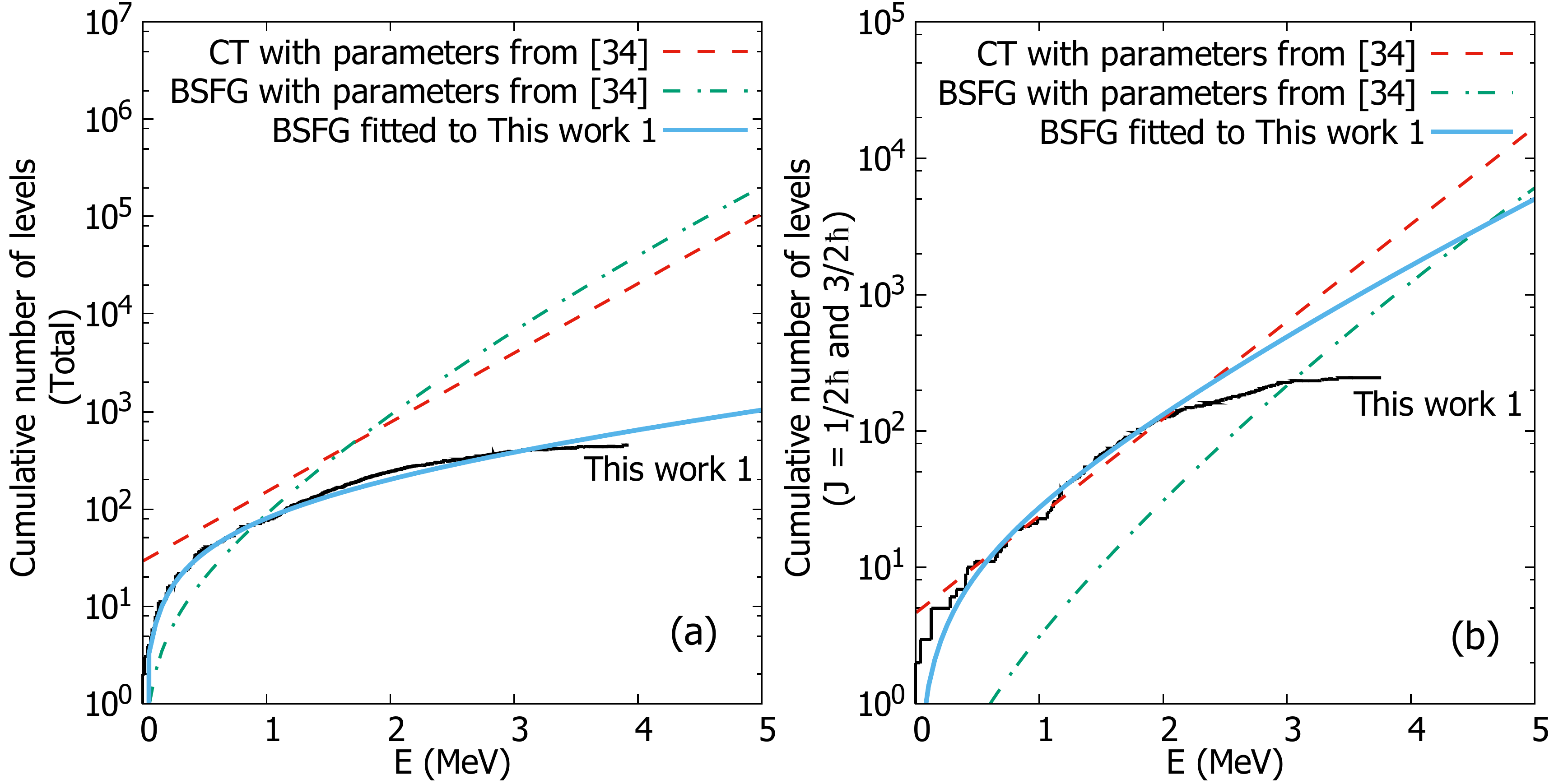}
\caption{\label{culm} (Color online) Comparison between the experimental total (a) and partial (b) cumulative numbers of levels and those predicted by two phenomenological NLD models.}
\end{figure}

The cumulative number of levels is very helpful for verifying the predictive power of the NLD models. In Fig. \ref{culm}, we compare our experimental cumulative curve (This work 1) with two phenomenological NLD models, namely the back-shifted Fermi gas (BSFG) and constant temperature (CT). The functional forms of these two models are taken from Ref. \cite{egidy1988}, that is
\begin{eqnarray}
&&\rho_{CT}(E,J)=f(J)\rho_{CT}(E)=f(J)\frac{1}{T}e^{(E-E_0)/T}~, \label{ct} \\
&&\rho_{BSFG}(E,J)=f(J)\rho_{BSFG}(E)=f(J) \frac{e^{2\sqrt{a(E-E_1)}}}{12\sqrt2\sigma a^{1/4}(E-E_1)^{5/4}}~, \label{fg} \\
&&f(J)=e^{-J^2/2\sigma^2} - e^{-(J+1)^2/2\sigma^2} \simeq \frac{2J+1}{2\sigma^2}e^{-(J+\frac{1}{2})/2\sigma^2}~, \label{fj}
\end{eqnarray}
where $\sigma_{CT}=0.98A^{0.29}$ and $\sigma_{BSFG} = 0.0146A^{5/3}\frac{1+\sqrt{1+4a(E-E1)}}{2a}$ are the spin cut-off parameters with $E_1$ and $a$ being the back-shifted energy and level density parameters, respectively. Two parameters $E_0$ and $T$ in Eq. (\ref{ct}) are the energy shift and constant temperature, whereas the function $f(J)$ in Eq. (\ref{fj}) is the conventional spin distribution of the NLD \cite{gcmodel}. The free parameters $a, E_1, E_0$, and $T$ of the BSFG and CT are often adjusted to fit the total cumulative number of levels as well as the NLD determined from the experimentally averaged neutron-resonance spacing data ($D_{0}$ value) \cite{egidy2005}. The values of these free parameters taken from Ref. \cite{egidy2005} (see also Table \ref{tab3}) were used to calculate $\rho_{CT}(E)$, $\rho_{BSFG}(E)$, $\rho_{CT}(E,J)$, and $\rho_{BSFG}(E,J)$ ($J=$[$\frac{1}{2},\frac{3}{2}$]$\hbar$). The total and $J$-dependent cumulative numbers of levels are then calculated making use of Eq. (\ref{ne_total}). The results obtained shown in Fig. \ref{culm}(b) indicate that the CT model with parameters taken from Ref. \cite{egidy2005} fits well to our experimental data (This work 1) for the spin range of [$\frac{1}{2},\frac{3}{2}$]$\hbar$, but it is higher than our experimental total cumulative curve (Fig. \ref{culm}(a)). The reason is that the parameters of the CT model taken from Ref. \cite{egidy2005} were given based on the analysis of 21 excited levels below 0.49 MeV within the spin range of [$\frac{1}{2},\frac{9}{2}$]$\hbar$ (close to the spin range of [$\frac{1}{2},\frac{3}{2}$]$\hbar$ within the present work), whereas below 0.49 MeV, there must be in total 37 excited levels within a much larger spin range of [$\frac{1}{2},\frac{19}{2}$]$\hbar$ as in the ENSDF library \cite{Helmer2006}. Consequently, while the CT model describes well the experimental $J$-dependent cumulative curve, it is unable to describe the total one. For the BSFG model with the free parameters taken from the same Ref. \cite{egidy2005}, it completely fails to describe both the total and $J$-dependent experimental cumulative curves (see Fig. \ref{culm}). The above results of the CT and BSFG models clearly demonstrate that the prediction of the phenomenological NLD models depends strongly on the values of their free parameters. For instance, by re-fitting the results of the BSFG model to our total and $J$-dependent experimental cumulative data, we obtain the different sets of free parameters as reported in Table \ref{tab3}. To obtain a reliable predicting power, one should therefore use the microscopic NLD models instead of the phenomenological ones.

\begin{table}[h]
\caption{Values of the free parameters obtained within the CT and BSFG models presented in Fig. \ref{culm}.}
\begin{tabular}{p{6cm}|c|c|c|c}
Model & \multicolumn{2}{c|}{CT} & \multicolumn{2}{c}{BSFG} \\ \hline
Parameter & $E_0$ (MeV) & $T$ (MeV) & $a$ (MeV$^{-1}$) & $E_1$ (MeV) \\ \hline
Parameters from \cite{egidy2005}& $-2.06 \pm 0.29 $ & $0.61 \pm 0.03$ & $17.76 \pm 0.28$ & $-1.08 \pm     0.13$ \\
Fitted to This work 1 in Fig. \ref{culm}(a)& - & - & $3.51 \pm 0.28$ & $-12.09 \pm 1.24 $ \\
Fitted to This work 1 in Fig. \ref{culm}(b)& - & - & $12.73 \pm 0.16$ & $-3.49 \pm 0.07 $ \\
\end{tabular}
\label{tab3}
\end{table}

\begin{figure}[h]
\includegraphics[width=8.6cm]{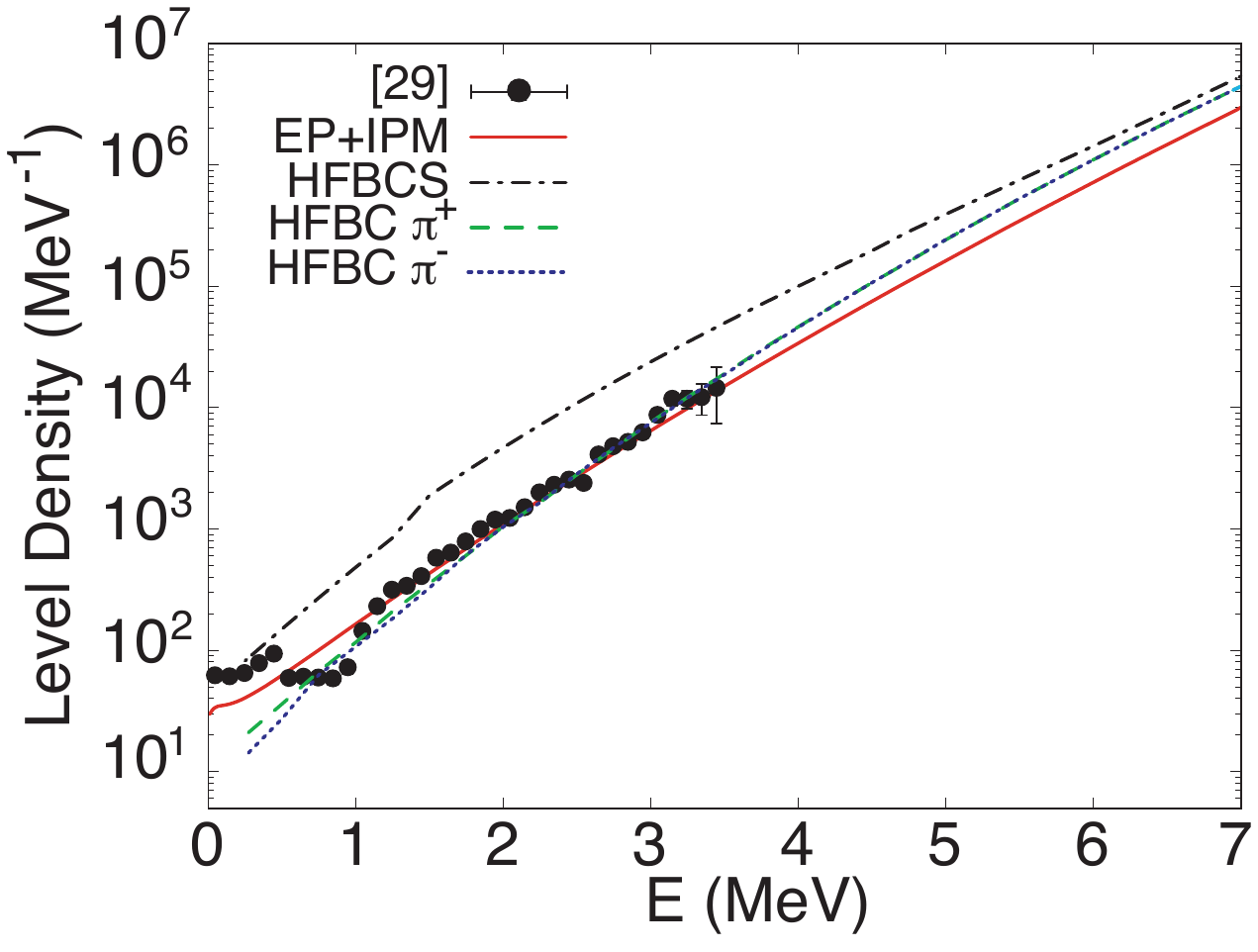} 
\caption{\label{nld1} (Color online) Comparison between the total NLDs obtained within different microscopic NLD models and the experimental data taken from Ref. \cite{Oslo}.}
\end{figure}

\begin{figure}[h]
\includegraphics[width=17.2cm]{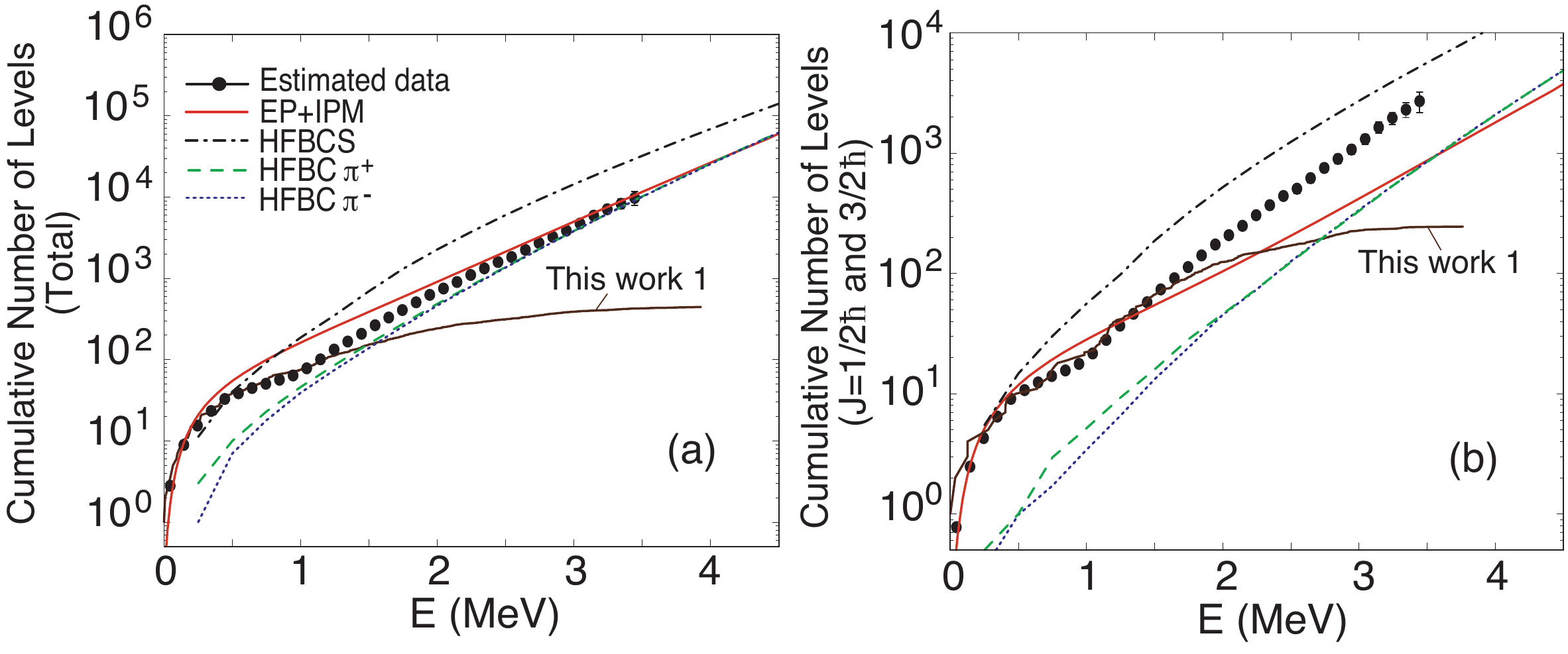} 
\caption{\label{culm1} (Color online) Total (a) and partial (b) cumulative numbers of levels obtained within different microscopic NLD models in comparison with the experimental data obtained within the present work (This work 1) and those calculated from the experimental NLD data in Ref. \cite{Oslo} (estimated data).}
\end{figure}

Within the present paper, three microscopic NLD models have been selected, namely the Hartree-Fock BCS (HFBCS) \cite{hfbcs}, the Hartree-Fock-Bogoliubov plus combinatorial method (HFBC) for the positive (HFBC $\pi^+$) and negative (HFBC $\pi^-$) parities \cite{hfb}, and the recent exact pairing plus independent-particle model at finite temperature (EP+IPM) \cite{epipm}. The HFBCS and HFBC data are accessible from RIPL-2 \cite{ripl2} and RIPL-3 \cite{ripl3}, respectively. These models have been considered to be the most up-to-date microscopic theoretical models for the NLD. Figure \ref{nld1} shows the total NLD $\rho(E)$ obtained within the HFBCS, HFBC, and EP+IPM in comparison with the experimental data. This figure indicates that while the HFBCS agrees with the experimental data only in the very low-energy region (below 0.5 MeV), both the HFBC and EP+IPM offer a good fit to the measured data. Moreover, the HFBC can not describe the data below 0.5 MeV, whereas the EP+IPM, in general, agrees with both low- and high-energy data. Consequently, one can easily see in Fig. \ref{culm1} that only the EP+IPM can describe both the experimental total and partial cumulative curves. This result of EP+IPM does not go beyond our expectation because this model has successfully been used to describe the NLD data of not only hot $^{170-172}$Yb \cite{epipm} and $^{60-62}$Ni \cite{epipm1} nuclei but also several hot rotating $A \sim 200$ isotopes \cite{epipm2}. In addition, the EP+IPM does not use any fitting parameters as discussed in Refs. \cite{epipm,epipm2,epipm3}, whereas the HFBCS and HFBC often employ some fitting parameters (see e.g., Eqs. (17) and (18) of Ref. \cite{hfbcs} or Eq. (25) of Ref. \cite{hfb}) to the experimental total cumulative data at low energy and the $D_0$ value at energy $E = B_n$. The above results, once again, confirm the microscopic nature and universality of the EP+IPM NLD model proposed in Ref. \cite{epipm}. In other words, the presently updated data provide a good test for both phenomenological and microscopic NLD models.


\section{Conclusion}
The present paper studies the excited levels of $^{153}$Sm nucleus populated in the thermal neutron-capture reaction using the $\gamma-\gamma$ coincidence technique and high resolution HPGe detectors. The coincidence technique together with the highly enriched target for $^{152}$Sm isotope allow us to significantly eliminate the influence of $^{150}$Sm excited nucleus in the observed gamma spectrum. In addition, the statistics of the measured data are rather high within the framework of coincident measurements. As a result, we are able to detect many new energy levels and their corresponding gamma transitions, namely 74 primary gamma transitions, 61 intermediate levels, and 291 secondary transitions. The tentative spin value of 53 observed levels is found to be $\frac{3}{2}\hbar$, whereas the remain levels are tentatively adopted to be in the spin range of $[\frac{1}{2},\frac{3}{2}]\hbar$.

By combining the updated energy levels with those obtained from the ENSDF library, we have constructed the new total and partial (within the spin range of $[\frac{1}{2},\frac{3}{2}]\hbar$) cumulative numbers of levels and compared the obtained data with those calculated from the experimental NLD data extracted by using the Oslo method (estimated data) as well as the predictions of different phenomenological and microscopic NLD models. The good agreement between our new cumulative curves with the estimated data allows us to deduce the values of the maximum excitation energy $E_{\rm max}$, which is defined as the energy threshold below which most of the excited levels have been observed, to be extended to around 1.2 and 1.8 MeV for the total and partial (spins of $[\frac{1}{2},\frac{3}{2}]\hbar$) NLD data, respectively. These values of $E_{\rm max}$ are higher than the corresponding values obtained by using the data presently existed in the ENSDF library. Moreover, the newly constructed cumulative curves also agree well with the recent microscopic exact pairing plus independent-particle model at finite temperature in which no fitting parameter has been employed.

All the results obtained within the present work are important as they will provide the updated information on the nuclear level structure and make a step forward to the completed level schemes of excited compound nuclei.

\begin{acknowledgments}
N.N.A, N.X.H, P.D.K, H.H.T, N.Q.H acknowledge the support by the National Foundation for Science and Technology Development (NAFOSTED) of Vietnam through Grant No. 103.04-2017.323. They would also like to thank the Ministry of Science and Technology of Vietnam for the financial support through the project coded KC05.08/16-20. Sincerely thanks are given to Prof. Vuong Huu Tan (former Chairman of Vietnam Atomic Energy Institute) and Prof. Nguyen Nhi Dien (former Director of Dalat Nuclear Research Institute) for their important decisions and supports to implement the neutron beams at Dalat Nuclear Research Reactor, which have been continuously used for the nuclear structure study in Vietnam.

\end{acknowledgments}


\end{document}